\documentclass[12pt]{article}

\newcommand{\be}{\begin{equation}}
\newcommand{\ee}{\end{equation}}
\newcommand{\ba}{\begin{eqnarray}}
\newcommand{\ea}{\end{eqnarray}}

\begin{document}
\setlength{\baselineskip}{.5cm}
\renewcommand{\thefootnote}{\fnsymbol{footnote}}
\newcommand{\lp}{\left(}
\newcommand{\rp}{\right)}

\begin{center}
\centering{{\bf \Large Economic returns of research\,:\\
the Pareto law and its implications\\}
\vskip 1cm
Didier Sornette $^{1,2}$ and Daniel Zajdenweber $^3$\\
$^1$ Laboratoire de Physique de la Mati\`ere Condens\'ee\\
CNRS UMR6622 and Universit\'e de Nice-Sophia Antipolis, B.P. 71\\
06108 Nice Cedex 2, France\\
sornette@naxos.unice.fr\\
$^2$ Institute of Geophysics and Planetary Physics \\and
Department of Earth and Space Sciences\\
UCLA, Box 951567, Los Angeles, CA 90095-1567\\
$^3$ SEGMI, Universit\'e de Parix X - Nanterre\\
200, Avenue de la R\'epublique, F-92001 Nanterre Cedex, France
}
\end{center}

\vskip 1cm
{\bf 
At what level should government or companies support research? This complex multi-faceted
question encompasses such qualitative bonus as satisfying natural
human curiosity, the quest for knowledge and the impact on education and culture, 
but one of its most scrutinized component reduces to the assessment of economic
performance and wealth creation derived from research. 
Many studies report evidences of positive economic benefits derived from basic research
\cite{Martin,NAS}. In certain areas such as biotechnology, semi-conductor physics, 
optical communications \cite{Ehrenreich}, the impact of basic 
research is direct while, in other disciplines, the path from discovery to applications
is full of surprises. As a consequence, there are persistent uncertainties in the quantification of 
the exact economic returns of public expenditure on basic research. This gives little help
to policy makers trying to determine what should be the level of funding.
Here, we suggest that these uncertainties have a 
fundamental origin to be found in the interplay between
the intrinsic ``fat tail'' power law nature of the distribution of economic returns, characterized by
a mathematically diverging variance, and the stochastic character of 
discovery rates. In the regime where the cumulative
economic wealth derived from research is expected to exhibit a long-term positive trend, we show 
that strong fluctuations blur out significantly the short-time scales\,:
a few major unpredictable innovations may provide a finite 
fraction of the total creation of wealth. In such a scenario, 
any attempt to assess the economic impact of research over a finite time horizon
encompassing only a small number of major discoveries is bound to be 
highly unreliable. New tools, developed in the theory of self-similar and complex systems
\cite{Dubrulleetal}
to tackle similar extreme fluctuations in Nature \cite{Mandelbrot},
can be adapted to measure the economic benefits of research, which is intimately associated to this
large variability.
}

\pagebreak 

\section{Introduction}

Basic research has provided enormous social public economic returns.
Striking examples can be put forward.
Modern communication is founded on fundamental research of 
electromagnetism and electron transport in semiconductors, which resulted in the transistor
and the derived electronics.
The laser used in medecine and many industrial applications 
resulted from basic research in optical pumping in atomic physics.
Mathematics is at the core of aircraft design, computing, prediction of climate change.
Global positioning system, which originated in the creation of atomic clocks for studying
relativity and quantum mechanics, has a wide range of applications (shipping, airlines...).
The Internet, which evolved from military and scientific computer networks, 
is one of the main component for the development of new information technologies,
which have grown to a \$500 billion industry. 

The case for increased government spending on research
rests on the assumption that basic research fuels
R\&D, which is the engine for a stronger economy. 
Whether this assumption is correct or not
has been debated for a long time, going back to Bacon who believed that technology
flows from academic science and to Adam Smith who maintained that it largely
derives from the industrial development of pre-existing technology \cite{Kealey}.
Technology is constantly evolving on its own and also in response to 
the progresses of basic science. Does
basic research confers a preferential economic advantage to countries and companies that
fund it \cite{Wong}? It has been argued that the accelerated path of technological
advances (for instance chips double in performance every 18 months) leads to an intense 
competition between companies that are more likely to rely on the high returns
that are obtainable from building products and services based on present knowledge rather
than on the unpredictable results of chancy basic research \cite{Odlyzko}. According to this view,
what matters is not creating new technology but absorbing and applying innovations 
quickly, because applying basic research to commercial products is long and expensive
and often produces unexpected results. Pushing these argument to the extreme, recall that,
almost a century ago in 1899, the head of the US Patent Office proposed to close up shop
because ``everything that can be invented has been invented''. In basic science, the 
anonymous peer review system is the gauge used to evaluate quality and to recommend
funding of researchers and projects. 
However, it is often said that C. Columbus would never have
left harbor if his voyage plans had been subjected to anonymous peer review. 
``Safe science'' and ``well-dressed'' trivia are negative side of the anonymous peer review
and of the publish-or-perish competition. In contrast, important innovations or discoveries
are extreme events much harder to fathom in advance and there are still many 
to be made. In his 1995 report, the president
of MIT, C. Vest, has listed our major ignorances, sorted out in the broad areas of mind, energy,
health, climate, space science, economy and information (see also Cazenave (1998) \cite{Cazenave}). 
For instance, we do not know
how we learn and memorize, how to synthetize new fuel for nuclear fission plants, how 
some genes mutate and lead to cancer; we do not know even in theory the degree of
predictability of climate, we do not know if other planets similar to ours can be found
in the Milky Way, why national economies evolve at different paces, what will be the impact of
global networks such as Internet on our societies.

Another approach is to imbed science in its social
context, suggesting an ``ecology'' of science in order to optimize adaptation to its
social, economic and technical environment \cite{Byerly}.
This is related to the developing field of ``industrial ecology'', which employs fully the analogy
between biological systems in a natural environment and industrial systems
designed and operated by humans. According to this analogy,
models of interactions between biological 
species are instructive to the study of the network of industrial 
processes, as the later involves also complicated interactions
such as the sharing of resources, the generation of the products and
the wastes. This study becomes vital for the society to maintain a
desirable carrying capacity, given continued economic, cultural, and 
technological evolution \cite{Graedel}. In ecology,
nonlinear interactions between species
often lead to a strongly intermittent ``punctuated'' dynamics with the potential for the
spontaneous appearence of catastrophic extinction events or bursts of
genetic diversity \cite{bak}. Cannot a similar behavior characterize scientific output?

\section{Proxy for the distribution of research economic benefits}

Measuring R\&D achievements is difficult, as most companies
seem not to keep these kinds of records and do not know what to say when asked what
outcomes are being realized from their R\&D investments \cite{Wolff}. Special 
benchmarking of different measures of R\&D performances and the impact of 
strategic management of technology are thus being developed \cite{Roberts}.
Already difficult as it is to appreciate the impact of R\&D investment in major companies,
the situation is worse for the quantification of the impact of basic science.
As a proxy for the distribution of incomes resulting from R\&D investment and basic
research, we propose to use data available from show business. Shocking as this
suggestion may seem, show business shares with research 
some of the main ingredients for success, such as talent, hard work, patience,
investment, modern technology such as computers and luck. And data is available. 
It is well-known that the artistic outputs are concentrated
among a few ``lucky'' individuals, leading to the ``superstar'' phenomenon, a not
uncommon observation also in the science community. For instance,
the fraction $s(i)$ of singers with $i$ gold-records for the period 1958-1989 is found to be
accurately described by the Yule distribution $s(i) = 1/i(i+1)$, which is
a power law with an exponent (defined as in (\ref{jfklkll}) below) equal to $\mu = 1$
\cite{Chung}. For the one hundred most successful performers, our own analysis 
indicates that the exponent increases to about $\mu = 2.7 \pm 0.1$. 

Another data set, more relevant to 
the question of the distribution of incomes resulting from investments in research, is the
distribution of earnings from
the most successful pictures in the movie industry in recent years. Similarly
to investment decision-making in R\&D and research, in order to approve a budget, 
studio executives have to make a judgment that
there is a sensible relationship between the cost of the film and
its potential revenues. They look at the potential earnings of a movie 
from all sources: video, television, foreign territories, merchandising,
soundtrack and theme park rides. The costs include
fees and salaries to the talent-actors, directors, producers, writers, 
length of the shooting schedule, stunts (car chases, crashes, airplanes, exploding buildings, 
fires), special effects on computers, studio overhead, etc. The success of a movie
in terms of its gross revenue is not always very predictable (viz. Waterworld)
and can vary in large proportions, as figure 1 illustrates. Figure 1a plots
the world wide gross revenue
from the theatres of the top box office 100 for year 1993 compiled on 3rd January 1994
by the trade newspaper ``Variety''. Amounts listed here reflect actual 
amounts received by the distributors, with
estimates made in the case of recent releases. 
Ideally, one should aggregate theatre revenue and video rental
income, as video rental has grown tremendously in the past years and
totals about half the total revenues.  However,  video rental is spread over a 
relatively long time period, in contrast to theatres for which the
data are known during the year following the release (the income is concentrated 
over a short period of time).  For simplicity, we thus only analyze the theatre
income. The cumulative distribution is represented with inversed axis, corresponding
to a so-called ``rank-ordering'' analysis, showing the $n$th
picture income $W_n$ as a function of the rank $n$. The first rank is Jurassic Park totaling
a revenue of more than $\$868$ millions, the second rank is The Fugitive totaling
$\$349$ million and so on. The double logarithmic axis qualifies a power law
distribution when the data aligns along a straight line\,:
\be
P(W) dW = {\mu \over (W/W_{min})^{1 + \mu}}~{dW \over W_{min}},
~~{\rm for}~~W_{min} \leq W < +\infty~~~~
{\rm with}~~\mu = 1.3 \pm 0.1~.
\label{jfklkll}
\ee
The crosses and squares
represent the dispersion values occurring with a probability equal to a half 
of the maximum likelihood, leading to $W_n~[1 \pm 1/\sqrt{\mu(n\mu +1)}]$
\cite{rankordering}. The exponent $\mu$ in (\ref{jfklkll}) is the inverse of the
slope of the fit in the rank-ordering plot.

This distribution (\ref{jfklkll}) 
is robust across different years. This is shown in figure 1b for years 1977 to 1994 for the 
 $20+$ biggest successes for each year. Data for 1993 and 1994 include worldwide income
 while previous years compile only the US and Canada revenues. 
 The exponent $\mu$ determined by two methods,
 a direct least-square fit of the rank-ordering plot and the Hill estimator \cite{Hill}, is shown
 for all the years from 1977 to 1994. The two measurements are consistent and provide
 an estimate of the error. All the data is consistent with a value of 
 $\mu \approx 1.5$ even if significant deviations from year to year can be observed.
 For $20$ points, the relative error in $\mu$ is about $25\%$. Note that, notwithstanding
 the change in accounting, $\mu$ remains robust at $1.5 \pm 0.3$. We further test
 this robustness by showing in figure 1d the
 rank ordering plot of the $20$ largest
ratios of gross revenue over budget for year 1993. The fit is of very good quality and
qualifies a power law with exponent $\mu \approx 1.55$.

The standard deviation for the $W$ variable is not defined for $\mu < 2$ 
(it is mathematically infinite), reflecting the fact that this power law distribution
(\ref{jfklkll}) has an extremely fat tail\,: for instance, in 1993, the first rank with
a revenue of more than $\$868$ millions is almost forty times larger than the $100$th rank with
a revenue of about $\$23$ millions!  It is remarkable that the exponent
$\mu \approx 1.5$ is very close to that of the distribution of wealth per capita 
in developed countries \cite{Levy}. The extrapolation to the impact of research 
of such power law distributions (\ref{jfklkll}) with a small exponent $\mu$ 
is compatible with the observation of a few exceptional case histories, for which
the economic benefits are enormous.

The existence of power law distributions in social phenomena has a long history 
(see \cite{Zajden} for a review) that
dates back at least to the social economist Pareto who found that the statistics of
income and the wealth distribution are described by a power law tail with exponent
$\mu \approx 1.5$ \cite{Pareto}. Closer to the productivity problem addressed here, Lokta
found that the percentage of authors publishing exactly $n$ papers as a function of $n$
is also a power law with $\mu \approx 1$ \cite{Lokta}.  More recently, 
Shochley analyzed in 1957 the scientific output of 88 research staff 
members of the Brookhaven National Laboratory in the USA. He found instead a log-normal 
distribution. Montroll and Shlesinger have shown that log-normal
distributions with large variance
can be mistaken for power laws over a quite large range \cite{Montroll}.
In the early sixties, Mandelbrot pointed out that stock market price variations 
are badly modelled by the Gaussian distribution and he proposed the use of L\'evy 
laws (with infinite variance) \cite{Mandel1,Mandel2}. 
Recent investigations show that the stock price variations
have finite variance and are more adequately described by truncated L\'evy laws \cite{mantegna} or 
stretched exponentials \cite{lahe}.

We now examine two implications of this power law distribution of revenues. 

\section{Research as an option in the decision process}

Decisions for investment are usually made using
conventional financial methods, using estimates of future cash flows. They fail
when applied to research and R\&D \cite{Newtonpearson}, because the problem is of 
a different nature. Research keeps open the option for later investment 
in production in new technology. 
It has been noticed that this problem can be formulated as an financial option problem\,:
a limited initial investment gives the investor the possibility but not the obligation
to invest further at the completion of the research in the production line.
This concept is implemented for instance in major pharmaceutical industries \cite{Merck} to help
decision in the suitability of the research on thousands of new molecules. Out of 
these, only a few will be developed and lead to a commercial success. 
Quantitatively, over the period 1965-1985, 
only 1787 new active substances have thus been introduced on the world market \cite{Huttin}

This approach in terms of options has been also advocated to cope with 
uncertainties in business, as a way to quantify the value and 
price of flexibility and adaptativity \cite{Mckinsey}. Take the
discovery by J. G. Bednorz and K.A. M\"uller of superconductivity in 
layered ceramic materials at a then-record-high temperature of 33 degrees above absolute zero.
This discovery set off an avalanche of research worldwide into related materials that yielded dozens
of new superconductors \cite{supra}, eventually reaching a transition temperature of 135 Kelvin.
Even among reknowned scientists, the conviction before this discovery
was that it was very unlikely that any breakthrough would occur
in superconductivity and beat the previous temperature barrier. This is an example
where keeping some flexibility in an apparent dead end paid off. Even if superconductivity
research does not seem very much profitable for a long time, it may pay to keep an option open.
A similar approach may be of value more generally for basic research.

Quantitative use of the option analogy to price R\&D have been used for instance in the 
the Pharmaceutical industry \cite{Merck}, within the canonical 
Black-Scholes-Merton option pricing model \cite{Hull}.
This model relies on
a view of the world uncertainties which use Gaussian distribution and the existence of 
a variance. A Gaussian distribution is characterized
by a mean and positive deviations from the mean larger than two standard deviations should not
occur more than $2.3~\%$ of the cases. Such distribution is completely unadapted to describe 
the huge range of impacts and potential benefits from rare breakthroughs or discoveries.
If we follow the model of revenue fluctuations suggested by eq.(\ref{jfklkll}), 
we see that the variance is theoretically infinite. In practice, this means that the 
estimation of the variance is strongly dependent on the specific finite realization
used to compute it. The variance fluctuates and increases as the size of the sample increases.
Thus, it cannot be used as an reliable estimation of the risk or uncertainty and 
Black-Scholes-Merton approach fails in this case. At present, there is no consensus
on a general theory that encompasses all cases but some progress has been made 
on the pricing and hedging of derivatives in the presence of 
power law distributions \cite{Bouchaudetal1,Bouchaudetal2}, that could be applied
to the R\&D pricing problem. A more general portfolio approach to research is required
since, in many cases, one has to deal with many options rather of a single one. Portfolio
optimization techniques have been developed in the presence of power law distributions
\cite{taminlarge}. New approaches are needed in the general case. 

The essence of the problem can be summarized by the Lindy effect \cite{Mandelbrot}\,: since
the expectation $\langle W \rangle|_{W > W_0}$ conditionned on events larger than $W_0$
is $\mu (\mu -1) W_0$ (for $\mu > 1$), this means that 
the future is proportional to the past! Mandelbrot vividly illustrated
the Lindy effect by the quote ``the future career expectation of a television comedian 
is proportional to his past exposure'' or with the parable of the young poets' cemetery
in which ``Anyone who stops young stops in the middle of 
a promising career'' (exact for $\mu = 2$). Such statements apply to researchers
and discoverers.

Let us finally stress that, in addition to the fat tail problem, 
we deal here with economic phenomena that are not well
arbitraged by a market process as in financial markets. Information is spread over many disparate
agents and is difficult to aggregate in a liquid market price process. Thus, the valuation of R\&D
options is in this sense closer to insurance claims for disasters (in inverse scale!) \cite{zajins}
than to financial derivatives.

\section{The intermittent nature of accrued research economic benefits}

Consider now the decision problem facing a nation or an international company on its
degree of commitment to research funding. If the revenues from research 
were deterministically predictable with small
fluctuations and with an obvious dependence on investment, the equation would be simple. The 
problem is that research profitability on the short term is highly unpredictable and
exhibits strong intermittency. 

What should be the annual level of research funding $F$ in order to maximize the
welfare of a nation? To address this question within a quantitative approach, we need to
specify the distribution of revenues derived from research and the impact of
investment on this distribution.

\subsection{The distribution of annual revenues}

Let us assume that the large fluctuations of returns from
a given R\&D investment are modeled
by the distribution (\ref{jfklkll}) with the same exponent $\mu$. 
This model amounts to discount all future
cash flows and other benefits to the time at which the discovery was made. 
Thus, an accumulation of discoveries
over time translates into a sum of instantaneous discounted cash flows. This procedure
becomes problematic for discoveries whose cash flows have a very long lifetime by 
bringing fundamental changes in the economy and in the style and quality of life
(electricity, transistors, antibiotics, etc).
In this sense, using the distribution (\ref{jfklkll}) may be conservative as the 
true distribution might have an even longer tail, i.e. an even smaller exponent $\mu$.

Budgets are usually prepared on a yearly basis. For accounting purpose, we thus need
to obtain the distribution of the total return from R\&D investments in a given year. 
A R\&D investment
made at time $0$ may lead to a breakthrough at time $1$ or later in the future if
funding continues. If the breakthrough is made at time $1$ after the investment is
made at time $0$, the 
return derived from it is discounted over all future cash flows derived from it and is
attributed to
this time period $1$. If no breakthrough is made, this 
is simply counted as a loss for the time period $1$. A discovery may take a long time and
require a long investment period. In this accounting scheme, the investments will be lost (in
reality they may prepare the next discovery)
until the year when the discovery is made at which all the future expected cashes flows
are discounted. Note that the procedure of 
counting as losses the investments that do not give fruit over
the next year does not imply that we a priori favor a short-term investment strategy. The 
potential importance
of long-term investment is implicitely taken into account into the ``fat tail'' power law
distribution (\ref{jfklkll}) of profits, i.e. in the (rare) occurrence of very large returns.

This addresses the question of the origin of very large returns. This would require a detailled
study on its own but let us suggest that very large returns for R\&D investment have probably
multiple inter-related sources, involving in particular luck and the product of accumulated efforts.
The power law (\ref{jfklkll}) would then result from at least two mechanisms and describe two kinds
of events\,: the first class are extreme events (lucky discoveries)\,; the second class
corresponds to breakthroughs that, while not entirely predictable, are made more probable by
a strong continuous commitment over long times. The magnitude of their profits, while still 
probably much larger than the cumulative investment, becomes commensurate with it. 

From our assumption that the distribution of returns from
a given R\&D investment is given by (\ref{jfklkll}), we obtain the distribution of
annual revenues due to research of a nation or a company. Since the annual revenue
is the sum of a possibly large number of contributions, the generalized central
limit theorem applies \cite{Taqqu}\,: in the limit of a very large number of 
contributions, the annual revenues are distributed according to a stable L\'evy
distribution with index equal to the exponent $\mu$. The L\'evy distribution is 
characterized by a power law tail of the same form as (\ref{jfklkll}). For a 
finite number of contributions, we simplify the representation of the distribution of annual revenues
by a simple powerlaw of the form (\ref{jfklkll}), with a value for $W_{min}$ normalized now to
represent an annual income. This simplified formulation is further justified by the fact that 
it is the only case that possesses the three properties of
1) stability under aggregation (sum of variables), 2) stability under mixing 
(of distributions) and 3) stability under choice of
extreme values \cite{zajbook}. Since the factors underlying the economic return of research
are many and complex, it is interesting that our empirical tests qualify the
distribution that is the most robust and adapted to these three relevant ingredients.

\subsection{Relationship between investment and distribution of revenues}

Consistent with the concept of universality
for self-similar systems \cite{Dubrulleetal}, we assume 
that the sole effect of changing the funding level
$F$ is to modify the minimum possible annual revenue $W_{min}$, while keeping the same
power law shape with the same exponent $\mu$ for the full distribution (\ref{jfklkll}) of potential
revenues derived from this funding effort. This assumption implies that the power law distribution
(\ref{jfklkll}) has a robust intrinsic origin rooted elsewhere than in the quantitative level
of investment, and which is to be found in self-organizing properties of social communities. 

The dependence of $W_{min}(F)$ 
is similar to that of production functions in neo-classical production
theory. One of the simplest such dependence assumes a homogeneous behavior
given by a generalization of the Cobb-Douglas function with constant elasticity 
$W_{min}(F) \sim L^a~F^{b-a}$, where $L$ is the labour quantity. For the application
to research, we assume full substitution between capital and research work force 
(most of the support goes to paying salaries and past investments are positively 
correlated with the quality and quantity of research labour) leading to a simple
functional dependence\,:
\be
W_{min}(F) = c ~F^b~,
\label{jklglg}
\ee
where $c$ is a generalized productivity (productivity is usually defined
as the ratio of output to input). We expect $0 < b \leq 1$, reflecting either
a self-similar behavior ($b=1$) or diminishing return rates ($b<1$).
Many other functional forms have been proposed which are qualitatively equivalent.
Expression (\ref{jklglg}) gives usually a good
approximation when optimum technicity holds
and represents correctly industries in which increase in size implies 
superposition of work force. 

Our last assumption is that funding is a fixed fraction $f$ of the gross national product $N_P$
\be
F = f ~N_P~.
\label{yjfdj,;w}
\ee
In the presence of correlations in the time series of profits (see below) and other 
economic factors, it may be favorable to have $f$ become a function of time. This leads
to an interesting optimization problem, left for another investigation.

\subsection{Resolution of the model}

We measure the welfare brought to the nation or company
by estimating its annual revenues. A more sophisticated approach involves
using more precise measures like utility functions, which we do not pursue here.
The average annual revenue of the nation or company is
\be
\langle W \rangle = \int_{W_{min}}^{+\infty} dW~W~P(W) = {\mu \over \mu - 1}~W_{min} \approx 4~W_{min},~~
~~~~~{\rm for}~~\mu = 1.3~.
\label{hsgfswx}
\ee

Starting from a gross national product $N_P(0)$ at initial time, the national 
product at time $n$ is
\be
N_P(n) = (1-f)~N_P(n-1) + v_{n-1}~c~(f ~N_P(n-1))^b~,
\label{ikkkkg}
\ee
if it was at level $N_P(n-1)$ the previous unit time. $v_{n-1}$ is a random number 
between $1$ and $+\infty$ drawn from the
normalized distribution $P(v) dv = \mu ~dv/ v^{1+\mu}$, such that 
$\langle v \rangle = {\mu \over \mu -1}$.
We have expressed $W_{min} = c~[f ~N_P(n-1)]^b$, as seen from (\ref{jklglg}) and (\ref{yjfdj,;w}).
The first term in the r.h.s. of (\ref{ikkkkg}) quantifies the cost of research funding.
The second term reflects the fluctuating nature of incomes resulting from research.

\subsubsection{$b=1$}

Consider the simplest case where 
wealth production from research is proportional to funding, i.e. $b=1$. Then, expression
(\ref{ikkkkg}) becomes 
\be
N_P(n) = (1- f + c~f~v_{n-1})~N_P(n-1)~,
\ee
which allows us to define the cumulative return $R(n)$ produced by the investment in research
\be
R(n) \equiv \ln {N_P(n) \over N_P(0)} = \sum_{i=0}^{n-1} \ln (1- f + c~f~v_i) 
\approx  \biggl( c~\sum_{i=0}^{n-1} v_i ~- ~n \biggl)~f~.
\label{jgkkgy}
\ee
The last approximate equality in (\ref{jgkkgy}) uses the fact that the funding 
and increase of gross national wealth are
tiny fraction (a few percent at most per year) of the total national product. 

On average, $c~\langle \sum_{i=0}^{n-1} v_n \rangle 
= c~n~[\mu/(\mu - 1)] \approx 4~c~n$ for $\mu = 1.3$, 
according to (\ref{hsgfswx}). Thus, the average return per unit time is
\be
R \equiv {1 \over n}~\langle R(n) \rangle = c~f~(4 - {1 \over c})~.
\label{jfjhsnxw}
\ee
If the generalized productivity $c$ of research is larger than $1/4$, 
the nation profits from research 
at the annualized return rate $cf(4 - 1/c)$. 
Take for instance $c=1/2$. This leads to an average yearly growth rate of the economy
exactly equal to funding ratio $f$.

Equation (\ref{jfjhsnxw}) shows that the average yearly return is proportional
to the funding level $f$ (by assumption (\ref{jklglg},\ref{yjfdj,;w}) for $b=1$)
and to the generalized productivity $c$. A sensible policy should thus strive
to increase productivity as the single most relevant factor in the presence of
budget constraints.

This is not the whole story\,: since the 
benefits of research are so wildly fluctuating according to their power law
distribution, the sum $\sum_{i=0}^{n-1} v_i$ is also distributed according to a distribution 
with a power law tail with the same exponent $\mu$ \cite{Taqqu,LevyPaul}.
This implies that the actual 
time evolution of the return $R(n)$ is a strongly fluctuating function of time. 

To get a better intuition of the intrinsic intermittent nature of economic returns from
research investment, we show in figure 2 a typical synthetic time series of 
the yearly economic growth rate  
$R(n) - R(n-1) = (c v_n - 1) f$ expressed in $\%$ as a function of time $n$ for $c=1/2$ and $f=1 \%$,
for a given realization of the random numbers $v_n$.
To make the presentation more suggestive, we present the time axis as corresponding
to the twentieth century.

The horizontal line at $1 \%$ is the average yearly growth rate. However, this average is
very rarely observed in a given year. It rather results from the fact that, most
of the time, the economic growth rate derived from research investment is slightly negative
but is puntuated by intermittent bursts of strong positive growths. 
The striking feature shown by figure 2 is that the economic growth
is mainly due to a few ``lucky'' discoveries.

Notice also the existence of apparent economic cycles in which recessions are preceded and
followed by strong growth periods. The sole ingredient that has been invoked to obtain
this phenomenology is the
power law distribution of annual returns. Short time series covering only a few decades
can thus give the misleading impression of order and of the existence of 
cycles while this may in fact result, as in
this example, from intermittent punctuated dynamics.
The point illustrated by these simulations
is that the benefit of research is very difficult to evaluate on short time
scales (of decades) if the wealth creation is indeed distributed with a very fat tail distribution.
This is the general property characterizing so-called L\'evy flights \cite{Levyflights},
of which the process $R(n)$ is an example.
If economists were to analyse the time series of figure 2, not knowing their power law
structure and using the standard (erroneous) assumption of Gaussian fluctuations, their 
econometric regressions would lead to completely unreliable estimations, because they would 
be strongly dependent on the specific time period used.  
What these simulations make clear is that, in presence of uncertain and rare but
dramatic discoveries, a funding policy made on short time scales is fundamentally 
ill-adapted to capture the intrinsic variability that produces 
the extraordinary potential of research on the long term. 

This intermittency
becomes even stronger when the productivity parameter $c$ decreases towards the 
threshold $1/4$. In contrast, the wealth created by research becomes more and more obvious 
as the productivity $c$ increases but $R(n) - R(n-1)$ and $R(n)$ still exhibit
the same large fluctuations.

Correlations can be easily introduced in the yearly returns $R(n) - R(n-1)$ so as to 
make the time series shown in figure 2 even more realistic, for instance by using convergent
multiplicative processes of the type first introduced in economy by Simon and Champenowne
to explain the growth laws for cities. Power laws like (\ref{jfklkll}) are
easily generated with additional interesting correlation structures \cite{kesten1} that
present similar structures to those of critical speculative markets \cite{Roehner}. We leave
their use in this context to another work.

Figure 3a presents a simulation covering ten thousand years of history. It
shows the cumulative return $R(n)/cf$ as a function of time 
$n$ for $c=1/3$, corresponding to 
a funding equal to $(4 - 1/c)/4=75\%$ of the average absolute research benefit, in other
words to a return equal to $4/3$ of the investment on average. This long time period
allows us to clearly identify the average trend given by 
$R \equiv = {R(n) \over n} = c~f~(4 - {1 \over c}) = {f \over 3}$ for $c=1/3$, as given by
(\ref{jfjhsnxw}). Again, the striking feature shown by figure 3 is that the economic growth
is mainly due to a few ``lucky'' discoveries, while the cumulative return may be even decreasing
over other long period of times as represented in figure 3b, showing that
there can be persistent times of apparently unproductive funding. 
As a consequence, 
research investments can be shouldered mainly by countries and major companies
which are robust to adverse fluctuations.

\subsubsection{$b<1$}

For a decreasing return rate $b<1$, the analysis is slightly modified. Taking
the expectation of (\ref{ikkkkg}), we get 
\be
\langle N_P(n) \rangle  = (1-f)~\langle N_P(n-1) \rangle + {\mu~c~f^b \over \mu-1} ~
\langle [N_P(n-1)]^b \rangle~.
\label{ikkkkqqg}
\ee
We consider a finite time interval over which $N_P(n)$ can be approximated as
distributed according to a power law distribution with exponent $\mu$, according to the law
of addition of power law variables \cite{LevyPaul}. This approximation amounts
to neglecting the difference between $\log (1+x)$ and $x$. Then, we can use the
relationship $\langle N_P(n-1))^b \rangle = {\mu - 1 \over \mu -b}~[N_{P~min}]^{b-1}~
\langle N_P(n-1) \rangle$ to get the average return per unit time
\be
R \equiv \ln {\langle N_P(n) \rangle \over \langle N_P(n-1) \rangle} \approx {\mu~c~f^b \over \mu -b}
~[N_{P~min}]^{b-1} ~-~ f~,
\label{fdjkgk}
\ee
which recovers (\ref{jfjhsnxw}) for $b=1$.

For $b<1$, $R$ increases for small $f$ due to the dominance of the first term in the r.h.s. of
(\ref{fdjkgk}) and decreases for large $f$ as the last term $-f$ takes over. There is thus an optimal
funding level 
\be
f^* = \biggl({\mu~c~b \over \mu -b}\biggl)^{1/(1-b)} ~[N_{P~min}]^{-1}
\ee
 for which $R$ is
maximum. Notice that  $f^*$ is a decreasing function of the total wealth.
Otherwise, the previous discussions on the importance of increasing the generalized
productivity and on the role of fluctuations still hold.

\subsection{Case $\mu < 1$}

One cannot rule out the possibility that the exponent $\mu$ of
the distribution of creation of wealth by research is less than one. This corresponds
to an even more dramatic situation since then the average gain per unit time
$\langle W \rangle$ becomes infinite mathematically as seen from (\ref{hsgfswx}). 
In practice, this means that the 
total {\it cumulative} return $R(n)$ given by (\ref{jgkkgy}) is completely controlled by the few
largest returns derived from a few discoveries in the whole time series. Quantitatively, for
instance for
$\mu = 2/3$, independently of the length of time over which the calculation is made, 
the largest revenue from a single discovery accounts typically for about $1/5$ of the 
total {\it cumulative} wealth creation over the whole history! This might be interpreted
as the impact of a new wide-ranging technology, 
such as electricity, that fundamentally modify future industries.
This regime is even harder to handle for policy makers since research funding
is most of the time unproductive as an open option, which may suddenly burst in an
extraordinary discovery. What
technologies of the future are being stunted by well-intentioned
efforts to curtail curiosity-driven research?

\section{Fluctuating discovery rates}

Up to now, we have aggregated all sources of fluctuations in the annual distribution
(\ref{jfklkll}) of income. This approximation amounts to neglect the dispersion in the
number and size of discoveries occuring during a given year. Let us now reintroduce this phenomenon.
We thus consider simultaneously two sources of fluctuations\,: (1) the number $k$ of
discoveries per year is fluctuating according to a distribution $p(k)$\,; (2) each
discovery produces a discounted income $w$ distributed according to a power law $P_w(w)$
distribution similar to (\ref{jfklkll}) with $W_{min}$ replaced by $w_{min}$. We
consider first the average yearly return and then the simple memoryless Poisson rate for
discoveries. In absence of precise
constraints on the rate of discoveries, we then investigate
the impact of a power law rate and long-range time
correlations in the discovery rate upon economic returns. This analysis underlines the importance
of characterizing the factors (possibly different) affecting both the discovery rate and the
size distribution of returns.

\subsection{Average yearly return}

The total return in a given year is the sum of the returns from all discoveries
made in this year and reads on average
\be
\langle W \rangle = \lambda ~\langle w \rangle = \lambda~ {\mu \over \mu-1} w_{min}~,
\label{ghjjss}
\ee
where $\lambda$ is the average number of yearly discoveries.
The value of $w_{min}$ is a function of extrinsic (perception threshold, significance, fixed
costs,...) and intrinsic (strategy, funding, threshold of the Pareto law, etc) parameters. Note that
$\lambda$ is also a function of the parameters determining $w_{min}$. It is 
an increasing function of $w_{min}$ for small $w_{min}$ (more funding leads to a larger
effort and a probably larger probability for a discovery) and
decreasing for large $w_{min}$ (as the threshold of significant discoveries increases, 
their rate decreases). Future investigations need to establish the relationship between
$w_{min}$ and $\lambda$ and the positive and negative feedback effects that result in the
expression (\ref{ghjjss}) of $\langle W \rangle$.

\subsection{Fluctuations of yearly returns}

The fluctuations of the total yearly income $W$ are described by the distribution
\be
P_W(W) = \sum_{k=1}^{\infty} p(k)~P_w^{\otimes k}(W)~,
\label{gdhjkskw}
\ee
where the symbol $P_w^{\otimes k}$ indicates that $P_w(w)$ has been convoluted $k$ times
with itself. This sum weights the different possible outcomes of the number $k$ of
discoveries per year whose cumulative returns sum up to $W$. 

\subsubsection{Poisson rate}

If discoveries are independent random events without memories or correlations, 
the distribution $p(k)$ is given by the Poisson law
\be
p(k) = e^{-\lambda}~{\lambda^k \over k!}~,
\ee
where $\lambda = \langle k \rangle$ is the average number of yearly discoveries. It is 
also the standard deviation $[\langle k^2 \rangle - \langle k \rangle^2]^{1/2}$.

The calculation of (\ref{gdhjkskw}) is easily performed by taking its Laplace transform and summing
the infinite series\,:
\be
{\hat P}_W(\beta) = \exp [\lambda ({\hat P}_w(\beta) -1)]~.
\label{gdhjdkkw}
\ee
 Since $P_w(w)$ is a power law
with exponent $\mu$, its Laplace transform is asymptotically (for small $\beta$
corresponding to large $w$ contributions) 
\be
{\hat P}_w(\beta) = 
\exp [- \gamma \beta - C |\beta|^{\mu}]~~~~~ ~{\rm for}~~ 1 < \mu < 2~ \cite{Physrevmoi}~,
\label{jjjdjjjjcj}
\ee
where $\gamma$ is proportional to the mean.
By expanding the exponential in (\ref{jjjdjjjjcj}) and putting it into (\ref{gdhjdkkw}), we get
\be
{\hat P}_W(\beta) \approx \exp [\lambda (- \gamma \beta - C |\beta|^{\mu}])]~,
\ee
showing that $P_W(W)$ is also a power law with the same exponent $\mu$ but with
a scale factor $W_{min}$ multiplied by $\lambda$.

\subsubsection{Power law distribution of discovery rate}

Let us consider an alternative extreme case in which the number $k$ of discoveries per year is
distributed according to 
\be
p(k) = {\nu \over k^{1+\nu}}~~~~~~{\rm for}~k\geq 1~.
\ee
The sum (\ref{gdhjkskw}) is more difficult to estimate exactly but its asymptotic 
expression is obtained by noting that its Laplace transform is of the form 
\be
{\hat P}_W(W) = \sum_{k=1}^{\infty} {\nu \over k^{1+\nu}}~e^{-[\ln {\hat P}_w(\beta)] k}
\approx \sum_{k=1}^{1 \over -\ln {\hat P}_w(\beta)} {\nu \over k^{1+\nu}} = 1 -
\biggl(-\ln {\hat P}_w(\beta)\biggl)^{\nu}~.
\ee
Using the expression (\ref{jjjdjjjjcj}), we get finally 
\be
{\hat P}_W(W) =  1 - \biggl( \gamma \beta + C |\beta|^{\mu}\biggl)^{\nu}~.
\label{fsggshsh}
\ee
For $\mu >1$, ${\hat P}_W(W) \approx 1 - \gamma^{\nu} |\beta|^{\nu}$
showing that $P_W(W) \sim \gamma^{\nu}/W^{1+\nu}$ is a power distribution with an 
exponent completely controlled by the fluctuation in the occurrence of discoveries.
For $\mu < 1$, the term $\gamma \beta$ is absent and 
$P_W(W) \sim C^{\nu}/W^{1+\nu \mu}$. In this case, both sources of fluctuations 
amplify the extreme character of the fluctuations.

\subsubsection{Long-range correlations between discoveries}

Let us assume that the correlation $C(t)$ between the number of discoveries in
two different years decays slowly with time as
\be
C(t) \equiv {\langle k(t) k(0) \rangle - \langle k(t) \rangle \langle k(0) \rangle
\over \langle k^2 \rangle - \langle k \rangle^2} \sim t^{-y}~~~~{\rm with}~0\leq y \leq 1~,
\ee
i.e. discoveries are correlated over long time scales.
The cumulative sum of returns over many years defines a fractional Brownian motion $B_H(t)$ with
fluctuations of typical amplitudes proportional to $t^H$, where the Hurst exponent is
given by $H= 1-{y \over 2}$ \cite{Feder}. 
We recover the usual Brownian random walk fluctuations for the
border case $y=1$ and for any correlation decaying faster.

Mathematically, Mandelbrot and Ness \cite{Ness} defined $B_H(t)$ as
\be
B_H(t) = {1 \over \Gamma(H+{1 \over 2})}~\int_{t_0}^t (t-t')^{H-{1 \over 2}} ~dW(t)~,
\ee
where $W(t)$ is the usual random walk (Wiener process) and $dW(t)$ is the infinitesimal
time increment of zero mean and variance equal to $dt$. This expression shows that, after
a long time after the initial investment performed at time $t_0$, the typical amplitude
of the fluctuations in the number of discoveries during the year $t$ is proportional to 
$(t-t_0)^{H-{1 \over 2}}$. Thus in this model, the longer the cumulative 
time over which investment in research is performed, 
the larger will the fluctuations be (as well as the average return)! Again, we find in this
scenerio that fluctuations are unavoidable.

\section{Concluding remarks}

This paper has attempted to provide a quantitative approach to the conundrum posed 
by the evaluation of the benefits and returns of research. 
Its motivation is rooted in the lively debate
blossoming in recent years within scientific and government agencies to address the decrease
of government funding and industrial R\&D investments. Instead of focusing on the search for
a solution to the question on the economic benefits of research, 
we have investigated what we believe is a necessary intermediate step before
reaching a full solution,
namely identifying the origin(s) of the difficulty. 
A first origin is methodological\,: the impact of research is often
fuzzy (spread out over a fraction of the society) and delayed in time. 
Indeed, important discoveries need
a suitable fertile background which derives from long-term investments in education and research and 
the aggregate cost entailled is very difficult to apportion to a set of discoveries. 
We have studied another 
source of uncertainty, stemming from the intrinsic variability of the discoveries, both in their
rate and in their importance, as well as in their derived returns.
Using returns from the Show Business as a proxy, 
we have shown that the distribution of returns is probably very wide, with the possibility
to observe very large events with a non-negligible probability. The concept of a typical
discovery or of a characteristic deviation from this typical value may become meaningless, 
since fluctuations dominate the process. The extraordinary large distribution of
potential benefits thus makes quantitative estimations unreliable if the methodology is
not carefully tailored to it. Standard econometric methods based on Gaussian assumptions
are bound to give unreliable and unstable results.
It is often stated that leading economists have estimated that technology has accounted 
for at least one-half of the
economic growth in advanced industrial nations in the last fifty years. If the wealth
derived from discoveries and innovation is indeed distributed according to a power law
such as (\ref{jfklkll}), this implies that any such estimate is very unstable and 
would demand a much longer time scale to be solidly based.

Instead of addressing the hard question of the economic return of research, a recent law,
the Government Performance and Results Act of 1993 in the USA \cite{Askingscience},
requires a related and somewhat simpler measure from its agencies, namely the
quantification of performance of investment in research with respect to pre-specified goals.
This approach is appropriate for the ``center'' of the distribution of benefits
but is completely inadequate for the unpredictable fat tail.
In view of the importance of the tail in the global balance, 
should not a cautious planning make room for
unpredictable ``extreme'' discoveries, i.e. find a subtle balance between the
optimization of the short-term research investment (the usual economic and politic point
of view) and the maturation over a long term of a favorable
environment for the flourishing of unpredictable new insights?

The present essay suggests to bring the problem of research economic benefits into 
the growing basket of natural and societal processes characterized by extreme behavior.
They range from large natural catastrophes such as volcanic
eruptions, hurricanes and tornadoes, landslides, avalanches, lightning strikes,
catastrophic events of environmental degradation,
to the failure of engineering structures, social
unrest leading to large-scale strikes and upheaval, economic drawdowns on national
and global scales, regional power blackouts, traffic gridlock, diseases and
epidemics, etc. These phenomena are extreme events that occur rarely,
albeit with extraordinary impact, and are thus completely under-sampled and thus 
poorly constrained. They seem to result from self-organising systems 
which develop similar patterns
over many scales, from the very small to the very large. There is an urgency
to assimilate in our culture and policy that we are embedded in extreme phenomena.
Our overall sense of continuity, safety and confort may just be an illusion stemming from
our myopic view. Let us unleash the battle of giants between extraordinary discoveries and
extreme catastrophes.

\vskip 1cm
A discussion with Nigel McFarlane in an early stage of this work is acknowledged.
We are grateful to L. Knopoff for a critical reading of a first version of the manuscript.

\pagebreak

\pagebreak

\noindent FIGURE CAPTIONS

\vskip 1cm

\noindent Figure 1a : Rank ordering plot of world wide gross revenues
from theatres of top hot box office 100 compiled on 3rd january 1994
by the journal ``Variety'' for the year 1993. 
Crosses and squares represent uncertainty intervals (see text).
\vskip 0.3cm
\noindent Figure 1b : Same as a) for the years 1977 to 1995 for the top 20 to 37 
(depending on the year). Year 1988 is not available. This data is 
compiled early january of the following year by the journal ``Variety''. The two 
straight lines corresponds to the best fits to year 1994 (top) and 1980 (bottom)
and have both a slope close to $2/3$ qualifying an exponent $\mu \approx 1.5$.
\vskip 0.3cm
\noindent Figure 1c : Variation of the exponent $\mu$ of the power law distribution
from 1977 to 1994, estimated by two methods\,: least square fit (thick line) and Hill
estimator (thin line). Both estimators give consistent results.
\vskip 0.3cm
\noindent Figure 1d : Rank ordering plot of the $20$ largest
ratios of gross revenue over budget for year 1993. Rank 1 corresponds to ``The wedding banquet''
with a return ratio of $23.6$\,: this movie had a small budget of $\$ 1$ million and gave
rise to a revenue $23.6$ times larger. The second rank is ``Jurassic Park'' with a return
ratio of $13.8$\,: it had a budget of $\$ 63$ million and gave
rise to a revenue $\$ 869$ millions.

\vskip 1cm

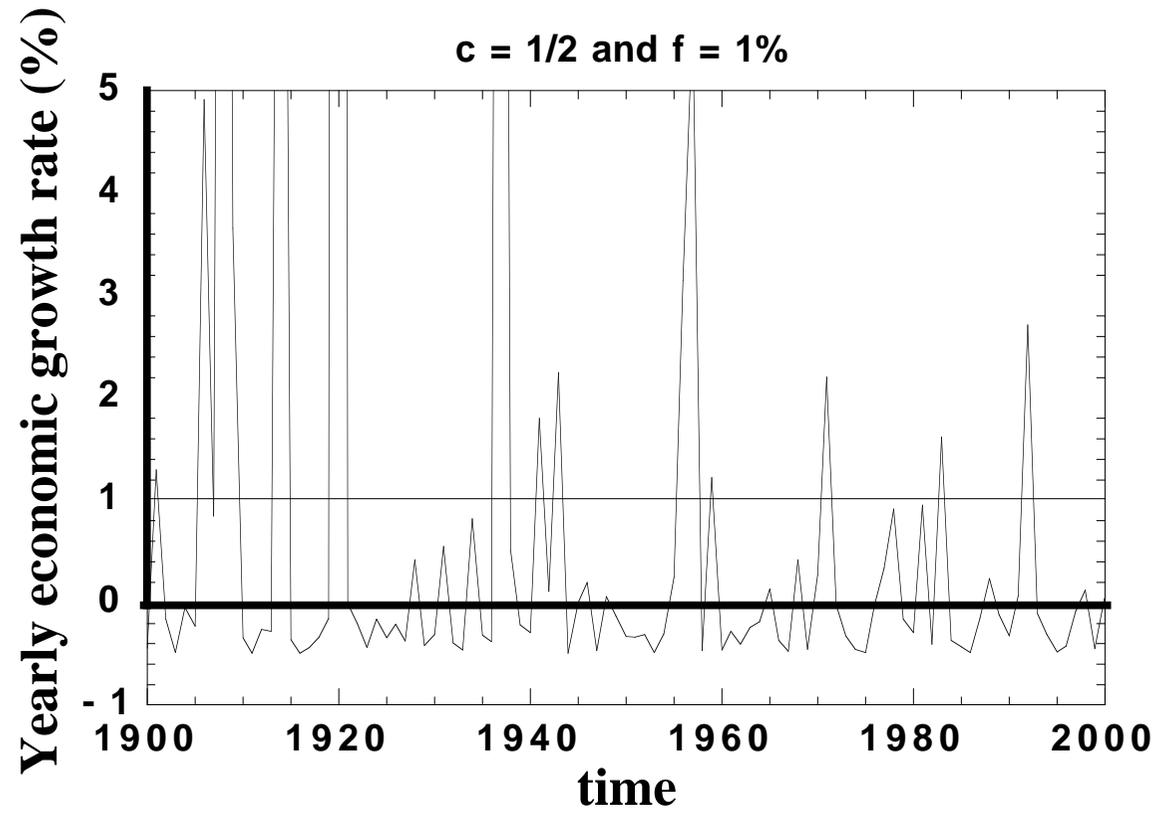
\noindent Figure 2 : A typical synthetic time series of 
the yearly economic growth rate  
$R(n) - R(n-1) = (c v_n - 1) f$ expressed in $\%$ as a function of time $n$ for $c=1/2$ and $f=1 \%$,
for a given realization of the random numbers $v_n$.
The horizontal line at $1 \%$ is the average yearly growth rate. 

\vskip 1cm

\noindent Figure 3 : a) Typical history of the cumulative return $R(n)/cf$, resulting from
research investment, as a function of time $n$ for a productivity $c=1/3$, corresponding to 
a funding equal to $(4 - 1/c)/4=75\%$ of the average absolute research benefit.

\noindent b) Part of the history shown in a).

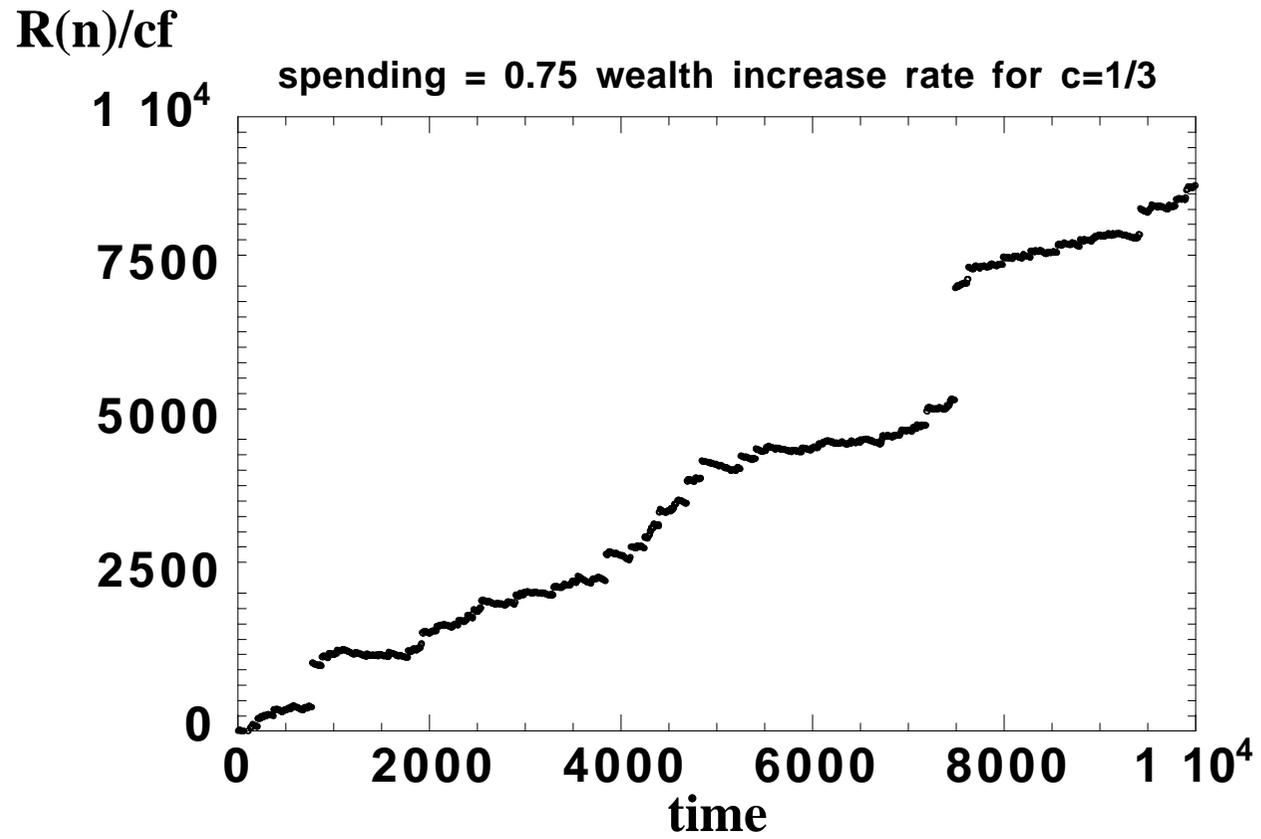
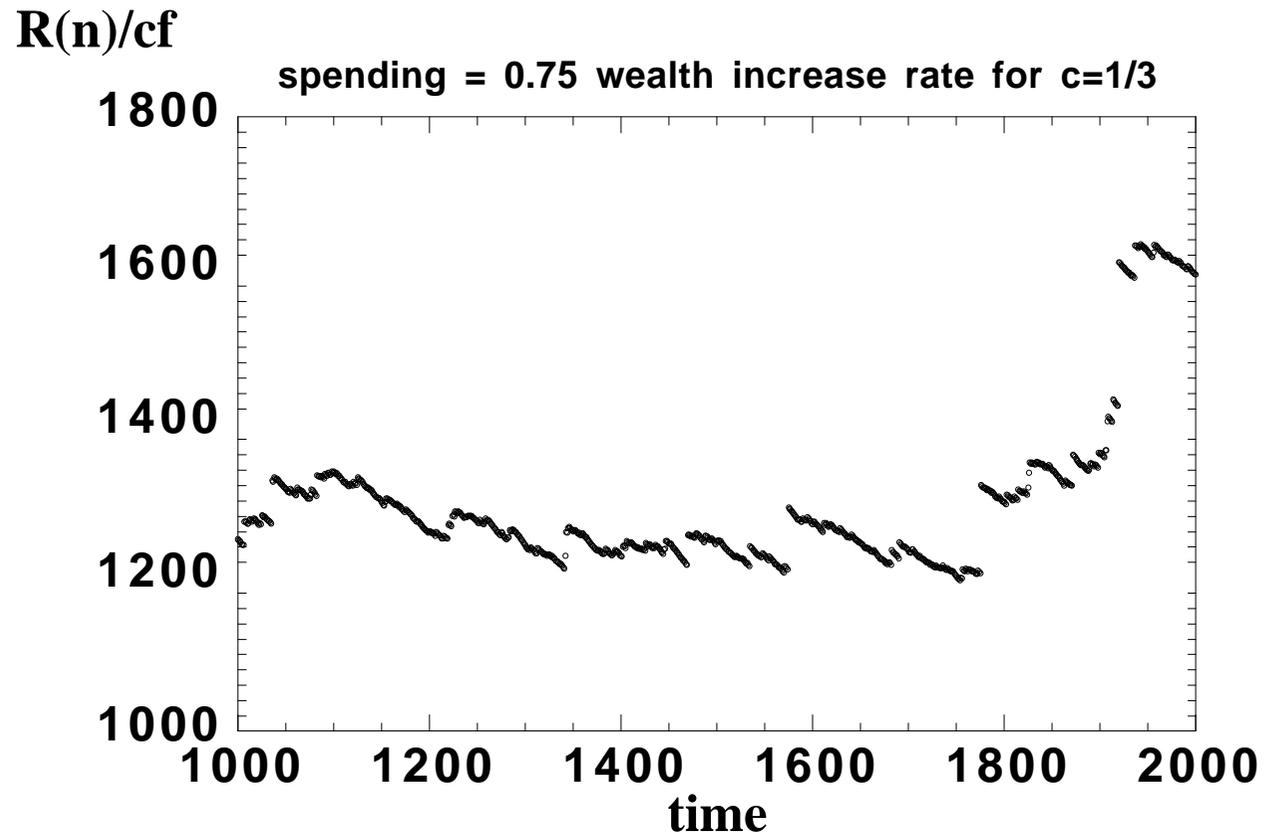
\end{document}